\documentclass[floatfix,aps, prd, reprint, amsfonts, amssymb, amsmath, showpacs, a4paper, nofootinbib, superscriptaddress]{revtex4-1}

\usepackage{color}
\usepackage{graphicx}
\usepackage{braket}
\usepackage{microtype} % To fix underfull hboxes
\usepackage{setspace} %% Fixes microtype errors

\newcommand{\be}{\begin{equation}}
\newcommand{\ee}{\end{equation}}
\newcommand{\bea}{\begin{eqnarray}}
\newcommand{\eea}{\end{eqnarray}}
\newcommand{\nn}{\nonumber\\}

\usepackage[caption=false]{subfig}

\begin{document}

\hfill{KCL-PH-TH/2025-31}

\title{Flavour interferometry in Reissner-Nordstr\"om background}

\author{Jean Alexandre}
\affiliation{Department of Physics, King's College London, London WC2R 2LS, United Kingdom}

\author{Emilio Meryn}
\affiliation{Department of Physics, King's College London, London WC2R 2LS, United Kingdom}

\begin{abstract}
    We derive the phase acquired by a neutral scalar particle propagating along Reissner-Nordstr\"om geodesics. Considering two flavours propagating on different trajectories which intersect, we plot the interference pattern induced by gravitational lensing from the charged compact object. 
    Although the effect of the charge is subdominant in the metric, it proves to be significant in the phase, and shifts the interference pattern, compared to the Schwarzschild case. 
    This pattern is characterised by two oscillation lengths which, if known, would allow the determination of both eigen masses independently.
\end{abstract}

\maketitle

\section{Introduction}

The rich topics of neutrino oscillation and gravitational lensing \cite{Virbhadra:1999nm} are mostly studied independently, but considered together lead to the field of neutrino optics \cite{Lambiase:2005gt}.
The effect of gravity on the phase of a quantum mechanical particle \cite{Stodolsky:1978ks} started with the original measurement on neutron phases in the Earth gravitational field \cite{Colella:1975dq},
and naturally led to the first studies of gravitationally-induced neutrino oscillations in the astrophysical context \cite{Fornengo:1996ef,Ahluwalia:1996ev}.
Works on particle phases in curved spacetimes were then developed and extended to different spins or astrophysical events \cite{Alsing:2000ji,Visinelli:2014xsa,Ramezan:2010zz,Sorge:2007zza,Yang:2017asl,Koutsoumbas:2019fkn,Swami:2020qdi,Swami:2021wbf,Dvornikov:2020oay}, but also in a context beyond General Relativity \cite{Chakrabarty:2021bpr,Chakrabarty:2023kld,Frost:2023enn,Alloqulov:2024sns,Lambiase:2020pkc,Hammad:2022azd,Wang:2025nyl}.

The interference pattern induced by flavour oscillations from lensing by a compact object was
studied in \cite{Alexandre:2018crg}, based on the original phase calculation in \cite{Fornengo:1996ef}. 
Assuming the compact object at the origin of coordinates, 
an interesting feature is the appearance of two oscillation lengths: {\it(i)} one in the radial direction, depending on the difference of masses squared;
{\it(ii)} the other in the orthoradial direction, depending on the sum of masses squared. 
Therefore the observation of such an interference pattern could in principle lead to the values of the eigen masses themselves.

In the present article we extend the study \cite{Alexandre:2018crg} to a Reissner-Nordstr\"om metric.
The radial propagation in this metric was already calculated in \cite{Huang:2004pd}, 
and we consider here the more general case of non-radial motion.
For a static, spherical and neutral compact object, the effects of spacetime curvature are at most of order $r_S/b$, 
where $r_S$ is the horizon radius and $b$ is the impact factor.
The effect of the electric charge, on the other hand, is felt at the order $(r_S/b)^2$ at most, similarly to the effect of the spin in Kerr metric.
As a consequence we naively do not expect a significant difference with the Schwarzschild case in the weak field approximation $b>>r_S$.
Nevertheless we find a significant difference in the interference pattern, between the Schwartzschild and the Reissner-Nordstr\"om case, because the effects of order $(r_S/b)^2$ accumulate as the particle propagates along a geodesic, and the resulting phase shift is not small.
We note that a thorough study of massless scalar wave scattered by a Reissner-Nordstr\"om black hole is done is \cite{Crispino:2009ki}, 
where the role of the charge also proves to be non-trivial.

Sec.\ref{sec:setting} introduces the general setting in which we derive the phase of a particle propagating in the Reissner-Nordstr\"om background.
In Sec.\ref{sec:phase} we evaluate this phase in the ultra-relativistic regime, 
and we propose a prescription to avoid artificial divergences and imaginary contributions, 
which appear when going to the weak-field approximation. We then plot the resulting interference patterns in Sec.\ref{sec:pattern},
which is significantly shifted, compared to the Schwarzschild case.

\section{General setting}\label{sec:setting}

\begin{figure}
    \centering
    \includegraphics[width=0.48\textwidth]{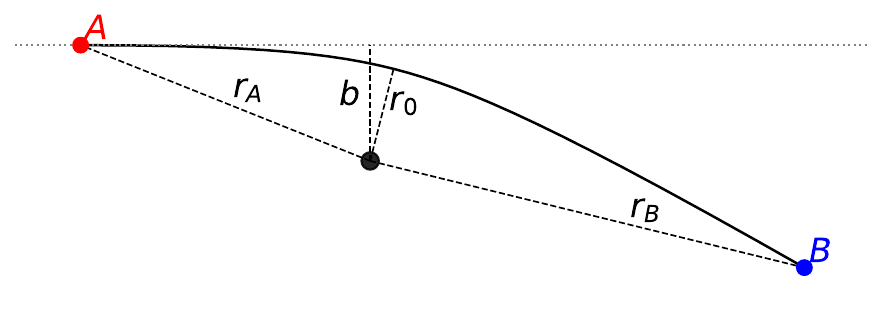}
    \caption{\label{scattering} The source is located at the point $A$, at the distance $r_A$ from the compact object, 
    and the detector is located at the point $B$, at the distance $r_B$ from the compact object. 
    $b$ is the impact parameter and the closest approach distance is denoted by $r_0$.}
\end{figure}

We consider a neutral scalar particle scattered by a charged compact object, 
and moving along a geodesic with impact parameter $b$ measured asymptotically far away, see Fig.\ref{scattering}.

The field corresponding to the particle is assumed to be a plane wave configuration.
The compact object is at the centre of coordinates $(r,\theta,\phi)$, and we focus on an equatorial motion (constant angle $\theta=\pi/2$), for which
the Reissner-Nordstr\"om metric is
\be\label{ds2}
ds^2=A(r)dt^2-\frac{dr^2}{A(r)}-r^2d\phi^2~,
\ee
where 
\be
A(r)=1-\frac{r_S}{r}+\frac{r_Q^2}{r^2}~.
\ee
The study is therefore reduced to a two-dimensional problem, or three-dimensional with cylindrical symmetry.
We consider $2r_Q\le r_S$, in which case the two horizons defined by $A(r)=0$ are located at
\be
r_\pm=\frac{1}{2}\left(r_S\pm\sqrt{r_S^2-4r_Q^2}\right)~.
\ee

The canonical momentum for a particle of mass $m$ is
\be
p_\mu=mg_{\mu\nu}\frac{dx^\nu}{ds}~,
\ee
and it satisfies the mass-shell relation.
\be\label{mass-shell}
m^2=g^{\mu\nu}p_\mu p_\nu=\frac{p_t^2}{A(r)}-A(r) p_r^2-\frac{p_\phi^2}{r^2}~.
\ee
The metric does not depend on $t$ or $\phi$, such that $E=p_t$ and $J=p_\phi$ are conserved quantities along the trajectory of a particle, 
and they correspond to the energy and angular momentum measured by an observer in asymptotically flat spacetime.

The phase acquired by a particle moving from $A$ to $B$ along a geodesic is
\be\label{phase}
\Phi=\int_A^B p_\mu dx^\mu=\int_A^B\left(Edt+p_rdr+Jd\phi\right)~,
\ee
and we will approximate the calculation of phase (\ref{phase}) for a null geodesic, which corresponds to the ultra-relativistic approximation,
relevant to the propagation of neutrinos. The energy and angular momentum which are constant along a null geodesic are denoted by $E_0$ and $J_0$ respectively. 
In the ultra-relativistic regime, we have then
\be
E=E_0+{\cal O}(m^2/E_0)~~~~\mbox{and}~~~~J_0=E_0b~.
\ee
It is known that the massless limit of geodesics followed by massive particles gives twice the phase obtained from null geodesics \cite{Lipkin:1995cb}.
This mismatch is related to the same-momentum approximation in the calculation of the phase, when assuming plane waves for the mass eigenstates \cite{Zhang:2003pn}, and can be solved by assuming wave packets instead. 
A related topic is the energy-time uncertainty principle in flavour oscillations, which has been studied in curved spacetime \cite{Blasone:2019jtj}.
The present article focuses on plane waves though, which as a first approximation is
enough to describe the interference pattern we are interested in.

\section{Phase along a geodesic}\label{sec:phase}

We follow here the steps given in \cite{Fornengo:1996ef} to obtain the generic form of the particle phase along a geodesic, 
but we consider an alternative evaluation of the integrals, as explained in Sec.\ref{sec:weak}.

\subsection{Radial motion}

We consider here a radial motion, for which $d\phi=0$, such that the phase along a null geodesic (denoted by the index 0) is
\be
\Phi=\int_{r_A}^{r_B}\left(\left.E\frac{dt}{dr}\right|_0+p_r\right)dr~,
\ee
and the mass-shell relation (\ref{mass-shell}) implies
\be
p_r=\pm\frac{\sqrt{E^2-A(r)m^2}}{A(r)}\simeq\pm\left(\frac{E}{A(r)}-\frac{m^2}{2E_0}\right)~,
\ee
where $\pm$ stands for inward or outward propagation. From the metric (\ref{ds2}), we also have for a radial motion,
\be
\left.\frac{dt}{dr}\right|_0=\mp\frac{1}{A(r)}~,
\ee
where the sign is opposite the one for $p_r$, such that the phase is 
\be\label{phaseradial}
\Phi=\frac{m^2}{2E_0}|r_B-r_A|~.
\ee
Note that the latter expression is identical to the one in flat spacetime, but here $|r_B-r_A|$ is not the 
proper distance $L$ over which particles propagate. The latter is instead
\bea\label{properlength}
L&=&\int_{r_A}^{r_B}\frac{dr}{\sqrt{A(r)}}\\
&=&\left[r\sqrt{A(r)} + \frac{r_S}{4} \ln\left(\frac{r\sqrt{A(r)}+r-r_S/2}{r\sqrt{A(r)}-r+r_S/2}\right)\right]_{r_A}^{r_B}~,\nonumber
\eea
and is larger than $|r_B-r_A|$. 

The calculation of the phase in a radial motion and in a Schwarzschild background is also done in \cite{Godunov:2009ce}.
There the momentum is expanded up to the order $m^4/E_0^3$, where $r_S$ appears explicitly in the expression for the phase.
This is not the case for the order $m^2/E_0$, where the effects of gravity would appear explicitly only if the phase (\ref{phaseradial}) was expressed in terms of the proper length (\ref{properlength}). As stated in \cite{Godunov:2009ce} though, the dependence on $r_S$ in the term of order $m^2/E_0$ is not specific to flavour oscillations, but arises only from the distinction between the proper length $L$ and the difference of coordinates $|r_B-r_A|$.

\subsection{Non-radial motion}

We first consider here a situation where the particle either approaches the compact object or recedes from it, 
therefore when the radial coordinate $r$ evolves monotonically in time.
The phase (\ref{phase}) can be written
\be
\Phi_{A\to B}=\int_{r_A}^{r_B}\left(E\left.\frac{dt}{dr}\right|_0+p_r+J\left.\frac{d\phi}{dr}\right|_0\right)dr~,
\ee
where
\bea
\frac{dt}{dr}&=&\frac{dt}{ds}\frac{ds}{dr}=-\frac{E}{A^2(r)p_r}\\
\frac{d\phi}{dr}&=&\frac{d\phi}{ds}\frac{ds}{dr}=\frac{J}{r^2A(r)p_r}~,
\eea
and $p_r$ is determined by the mass-shell relation (\ref{mass-shell}):
\bea\label{expandpr}
A(r)p_r&=&\pm E\sqrt{1-\frac{J^2A(r)}{r^2E^2}-\frac{m^2}{E^2}A(r)}\\
&\simeq&\pm E_0\sqrt{1-A(r)\frac{b^2}{r^2}}\mp\frac{m^2A(r)}{2E_0\sqrt{1-A(r)b^2/r^2}}~,\nonumber
\eea
where the null geodesic approximation is used. The expression for the phase simplifies as
\bea\label{phaseAB}
\Phi_{A\to B}&\simeq&\pm\int_{r_A}^{r_B}\left(\frac{E_0^2}{A^2(r)p_r}-p_r-\frac{b^2E_0^2}{r^2A(r)p_r}\right)dr\nn
&=&\pm\frac{m^2}{2E_0}\int_{r_A}^{r_B}\frac{dr}{\sqrt{1-A(r)b^2/r^2}}~,
\eea
where the signs - and + correspond to the particle approaching and receding, respectively. 

A comment is in order here: The expansion (\ref{expandpr}) is valid only if the argument of the square root does not vanish.
But it actually does vanish for one specific value of $r$, when the particle reaches the closest approach point $C$ along a full geodesic. 
Indeed, in this case, the radial coordinate is stationary at $r=r_0$ corresponding to $p_r(r_0)=0$.
For null geodesics this happens for
\be\label{defr0}
1=A(r_0)\frac{b^2}{r_0^2}~,
\ee
where the integrand appearing in the phase (\ref{phaseAB}) diverges, and the expansion (\ref{expandpr}) is not valid. 
The corresponding singularity is integrable though, since in the vicinity of $r_0$ the integrand is $\sim|r-r_0|^{-1/2}$.
As a consequence, the expression (\ref{phaseAB}) can be used in the integral defining the phase and, taking into account the $\pm$ signs for the approaching/receding branches, the phase acquired along a full (null) geodesic is 
\bea\label{PhiACB}
\Phi_{A\to C\to B}&=&\Phi_{A\to C}+\Phi_{C\to B}\\
&=&\frac{m^2}{2E_0}\int_{r_0}^{r_A}\frac{dr}{\sqrt{1-A(r)b^2/r^2}}\nn
&&~~~~+\frac{m^2}{2E_0}\int_{r_0}^{r_B}\frac{dr}{\sqrt{1-A(r)b^2/r^2}}~.\nonumber
\eea
We stress that this expression only assumes the ultra-relativistic regime, but at this stage does not involve any other approximation.

\subsection{Weak-field approximation}\label{sec:weak}

The integrals appearing in the phase (\ref{PhiACB}) cannot be evaluated exactly, and from now we assume a large-enough impact parameter,
satisfying $b\gg r_S > r_Q$. From its definition (\ref{defr0}), one can then express $r_0$ in terms of $b$ 
\be
r_0=b\left(1-\frac{r_S}{2b}-\frac{3r_S^2}{8b^2}+\frac{r_Q^2}{2b^2}+\cdots\right)~,
\ee
where dots represent higher orders in $r_S/b$ and $r_Q/b$. The integrand for the calculation of the phase (\ref{PhiACB}) can naively be expanded as
\bea\label{expansion}
&&\frac{1}{\sqrt{1-A(r)b^2/r^2}}\\
&=&\frac{1}{\sqrt{1-b^2/r^2}}-\frac{b^2r_S}{2(r^2-b^2)^{3/2}}\nn
&&+\frac{b^2r_Q^2}{2r(r^2-b^2)^{3/2}}+\frac{3b^4r_S^2}{8r(r^2-b^2)^{5/2}}+\cdots\nonumber
\eea
This expansion artificially introduces divergent integrals at $r=b$ though, as well as imaginary parts, since $r_0<b$. 
To deal with these artifacts, we introduce the following prescription. 
We analytically continue the expression (\ref{expansion}) to negative values of $r_S$, which leads to 
\be
r_0\to r_1=b\left(1+\frac{|r_S|}{2b}-\frac{3r_S^2}{8b^2}+\frac{r_Q^2}{2b^2}+\cdots\right)~>b~.
\ee
We then evaluate the (convergent) integrals, and go back to positive values of $r_S$. The resulting imaginary parts 
cancel out between the approaching and receding regimes, for the following reason.
In order to define the square root, one must introduce a branch cut, which can be chosen as the negative imaginary axis. 
When the particle approaches the closest point $C$, $z=r^2-b^2$ goes from positive to negative, corresponding to a multiplication by $e^{i\pi}$, such that the square root is multiplied by $e^{i\pi/2}=+i$. 
When the particle recedes from the closest point, $z$ goes from negative to positive, which corresponds to a multiplication by $e^{-i\pi}$ if one wants to avoid the  branch cut, such that the square root is multiplied by $e^{-i\pi/2}=-i$.
This approach reproduces the results given in \cite{Fornengo:1996ef} for the first order in $r_S/b$, and can easily be applied to higher orders, as shown below.

\subsubsection{Zero-th order in $r_S$}

The integral does not diverge at this order, and we have
\bea
\int_{r_0}^r\frac{du}{\sqrt{1-b^2/u^2}}&=&\sqrt{r^2-b^2}-\sqrt{r_0^2-b^2}\\
&=&\sqrt{r^2-b^2}\pm i\sqrt{|r_0^2-b^2|}~,\nonumber
\eea
such that, taking into account the cancellation of the imaginary parts,
\bea
&&\int_{r_0}^{r_A}\frac{dr}{\sqrt{1-b^2/r^2}}+\int_{r_0}^{r_B}\frac{dr}{\sqrt{1-b^2/r^2}}\\
&=&\sqrt{r_A^2-b^2}+\sqrt{r_B^2-b^2}.\nonumber
\eea
This result is intuitive: the zero-th order in $r_S$ corresponds to the absence of gravity, where a geodesic is a straight line and $r_0=b$.
The distance covered by the particle between $A$ and $B$ is then simply given by Pythagoras' theorem.

\subsubsection{First order in $r_S$}

The above prescription leads to
\bea
&&-\frac{b^2r_S}{2}\int_{r_0}^r\frac{du}{(u^2-b^2)^{3/2}}\\
&\to&+\frac{b^2|r_S|}{2}\int_{r_1}^r\frac{du}{(u^2-b^2)^{3/2}}\nn
&=&-\frac{|r_S|}{2}\left(\frac{r}{\sqrt{r^2-b^2}}-\frac{r_1}{\sqrt{r_1^2-b^2}}\right)~,\nonumber
\eea
and changing back the sign of $r_S$ gives
\be
\frac{r_S}{2}\left(\frac{r}{\sqrt{r^2-b^2}}\pm\frac{ir_0}{\sqrt{|r_0^2-b^2|}}\right)~.
\ee
With the cancellation of the imaginary parts we obtain
\bea\label{firstorder}
&&-\frac{b^2r_S}{2}\left(\int_{r_0}^{r_A}\frac{dr}{(r^2-b^2)^{3/2}}+\int_{r_0}^{r_B}\frac{dr}{(r^2-b^2)^{3/2}}\right)\nn
&\equiv&\frac{r_S}{2}\left(\frac{r_A}{\sqrt{r_A^2-b^2}}+\frac{r_B}{\sqrt{r_B^2-b^2}}\right)~,
\eea
where $\equiv$ means equal in the present prescription.
Note that this result is identical to the one derived in \cite{Fornengo:1996ef}, since it can also be written
\be
\frac{r_S}{2}\left(\frac{b}{\sqrt{r_A^2-b^2}}+\frac{b}{\sqrt{r_B^2-b^2}}
+\sqrt{\frac{r_A-b}{r_A+b}}+\sqrt{\frac{r_B-b}{r_B+b}}\right)~.\nonumber
\ee

\subsubsection{Second order in $r_Q$}

The procedure explained above leads to
\bea
&&\frac{b^2r_Q^2}{2}\int_{r_0}^r\frac{du}{u(u^2-b^2)^{3/2}}\\
&\to&-\frac{r_Q^2}{2}\left(\frac{1}{\sqrt{r^2-b^2}}+\frac{1}{b}\arctan\left(\sqrt{r^2/b^2-1}\right)\right)\nn
&&~~~~~~~~~~~~~\pm ~~\mbox{imaginary part}~,\nonumber
\eea
such that
\bea
&&\frac{b^2r_Q^2}{2}\left(\int_{r_0}^{r_A}\frac{du}{u(u^2-b^2)^{3/2}}+\int_{r_0}^{r_B}\frac{du}{u(u^2-b^2)^{3/2}}\right)\nn
&\equiv&-\frac{r_Q^2}{2}\left(\frac{1}{\sqrt{r_A^2-b^2}}+\frac{1}{b}\arctan\left(\sqrt{r_A^2/b^2-1}\right)\right)\nn
&&~~~~~~~~~~~~~~~~~+~(r_A\to r_B)~.
\eea

\subsubsection{Second order in $r_S$}

We have here 
\bea
&&\frac{3b^4r_S^2}{8}\int_{r_0}^r\frac{du}{u(u^2-b^2)^{5/2}}\\
&\to&\frac{3r_S^2}{8}\left(\frac{3r^2-4b^2}{3(r^2-b^2)^{3/2}}+\frac{1}{b}\arctan\left(\sqrt{r^2/b^2-1}\right)\right)\nn
&&~~~~~~~~~~~~~\pm ~~\mbox{imaginary part}~,\nonumber
\eea
such that
\bea
&&\frac{3b^4r_S^2}{8}\left(\int_{r_0}^{r_A}\frac{du}{u(u^2-b^2)^{5/2}}+\int_{r_0}^{r_B}\frac{du}{u(u^2-b^2)^{5/2}}\right)\nn
&\equiv&\frac{3r_S^2}{8}\left(\frac{3r_A^2-4b^2}{3(r_A^2-b^2)^{3/2}}+\frac{1}{b}\arctan\left(\sqrt{r_A^2/b^2-1}\right)\right)\nn
&&~~~~~~~~~~~~~~~~~+~(r_A\to r_B)~.
\eea

Finally, in the ultra-relativistic regime and weak-field approximation, the expression for the phase  is
\be\label{finalphase}
\Phi_{A\to C\to B}=\Psi(r_A)+\Psi(r_B)~,
\ee
where
\bea\label{Psi}
&&\Psi(r)=\frac{m^2}{2E_0}\Biggl\{\sqrt{r^2-b^2}+\frac{r_S}{2}\frac{r}{\sqrt{r^2-b^2}}\\
&&-\frac{r_Q^2}{2}\left(\frac{1}{\sqrt{r^2-b^2}}+\frac{1}{b}\arctan\left(\sqrt{r^2/b^2-1}\right)\right)\nn
&&+\frac{3r_S^2}{8}\left(\frac{3r^2-4b^2}{3(r^2-b^2)^{3/2}}+\frac{1}{b}\arctan\left(\sqrt{r^2/b^2-1}\right)\right)\Biggr\}~\nn
&&~~~~~~~~+{\cal O}(r_S/b)^3\nonumber
\eea
Note that the above expansion in $r_S/b$ is valid as long as its coefficients are of the same order at most. Comparing the zeroth and first order coefficients, $r_A$ and $r_B$ should then satisfy
\be
\frac{r_{A,B}(b/2)}{\sqrt{r_{A,B}^2-b^2}}\lesssim\sqrt{r_{A,B}^2-b^2}~~~\to~~~ r_{A,B}\gtrsim \frac{5}{4}b~,
\ee
in which case the second order coefficient is also of the same order. 
In the numerical analysis which follows, we will therefore assume that both $r_A$ and $r_B$ are larger than $5b/4$.

\section{Interference pattern}\label{sec:pattern}

\begin{figure}[htpb]
    \centering
    \includegraphics[width=0.48\textwidth]{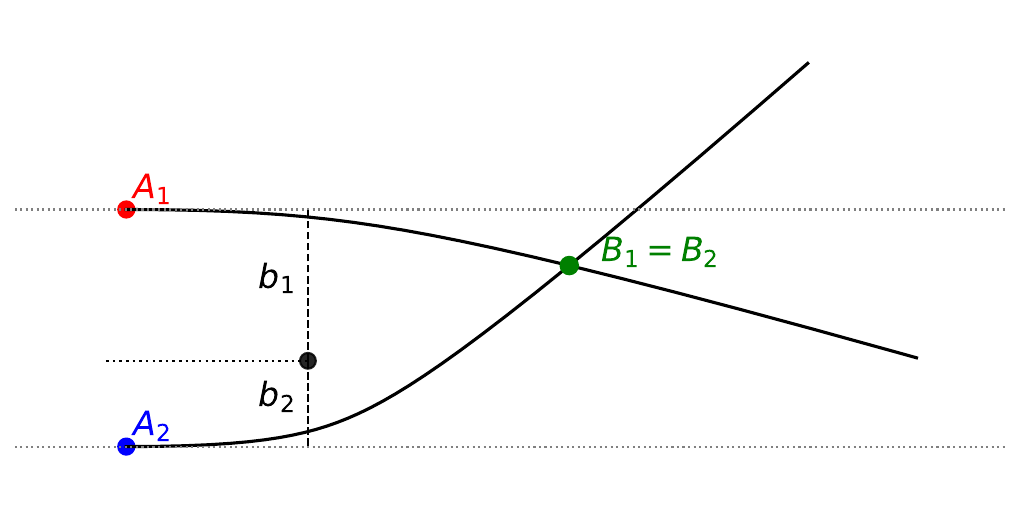}
    \caption{\label{interference} Four phases interfere at the intersection point after scattering: two flavours on each trajectory.}
\end{figure}

Having calculated the phase of one particle traveling along a geodesic, we can now study the interference between two coupled flavours on two different geodesics which intersect (see Fig.\ref{interference}). We assume here that the two emission points $A_1$ and $A_2$ are in phase. This corresponds for example to one specific astrophysical event emitting a coherent flux of coupled flavours, and is far enough to justify the plane wave approximation. We will estimate below how important this assumption is.

\subsection{Oscillation probability}

The two different trajectories are part of an incoming plane wave, which is scattered by the compact object. 
We therefore consider a source located far away from the compact object ($r_A\to\infty$)
in a way which does not affect the phase (\ref{finalphase}), by taking \cite{Alexandre:2018crg}
\be
\frac{m^2}{2E_0}\sqrt{r_A^2-b^2}\equiv 2q\pi\to\infty~,
\ee
where $q$ is an integer going to infinity. In the expression (\ref{finalphase}), the contribution from $r_A$ becomes then
\be\label{PsiA}
\Psi(r_A)\to 2q\pi+\frac{m^2}{4E_0}\left(r_S-\frac{\pi r_Q^2}{2b}+\frac{3\pi r_S^2}{8b}\right)~,
\ee
which corresponds to a non-trivial phase shift, independent of $r_A$. In the situation where $r_B$ also goes to infinity, this phase shift doubles, 
since the contributions of $r_A$ and $r_B$ are additive.
The phase shift is then independent of the distance between the source and the detector, potentially providing a measurable effect. 
As pointed out in \cite{Fornengo:1996ef}, the net phase shift induced when multiple gravitational sources are passed is then the sum of the phase shifts induced by the individual gravitational sources.

At the intersection point there are then four different phases to consider, 
denoted by $\Phi_i^J$, for two flavours $i$ and two trajectories $J$. Discarding the multiple of $2\pi$, we have 
\be\label{Phiij}
\Phi_i^J=\frac{m_i^2}{4E_0}\left(r_S-\frac{\pi r_Q^2}{2b_J}+\frac{3\pi r_S^2}{8b_J}\right)+\Psi_i^J(r_B)~,
\ee
where $\Psi_i^J(r)$ is given by eq.(\ref{Psi}), with $m\to m_i$ and $b\to b_J$. 
The flavour-changing oscillation probability was calculated in \cite{Fornengo:1996ef} and reads
\bea\label{proba}
P_{f\to f'}&=&\frac{\sin^2(2\Omega)}{8}\\
&\times&\Big(2+\cos(\Phi_1^1-\Phi_1^2)+\cos(\Phi_2^1-\Phi_2^2)\nn
&&~~~~-\cos(\Phi_1^1-\Phi_2^1)-\cos(\Phi_1^2-\Phi_2^2)\nn
&&~~~~-\cos(\Phi_1^1-\Phi_2^2)-\cos(\Phi_1^2-\Phi_2^1)\Big)~,\nonumber
\eea
where $\Omega$ is the mixing angle and $f,f'$ are the two flavours. 
The phases (\ref{Phiij}) are proportional to the eigen mass squared, such that the measurement of the probability (\ref{proba})
for different values of $r_B$ would lead to the values of $m_1^2,m_2^2$ independently, and not only the difference $m_1^2-m_2^2$.

Regarding the assumption of flavours being in phase at the emission points, one can estimate from the expression (\ref{PsiA}) what 
the change $\delta b$ in impact parameter should be, in order to compensate a random modification $\Psi\to\Psi+\alpha$ in the phase, with $\alpha\ll2\pi$. 
This compensation happens when 
\be
d\Psi=\frac{\partial\Psi}{\partial b}\delta b+\alpha=0~,
\ee
and therefore
\be
m\delta b=\frac{8\alpha}{\pi}~\frac{E_0}{m}~\frac{b^2}{3r_S^2/4-r_Q^2}~.
\ee
Given the regime we consider ($m\ll E_0$ and $r_S\ll b$), we have $|m\delta b|\gg\alpha$. 
Hence even a small phase variation could be equivalent to a large change in impact parameter, in units of $1/m$. 
As a result the interference pattern could easily be modified, and coherence of the incoming trajectories is essential for this study.

\subsection{Expansion for large distances}

We pointed out that our results are valid for regions in which $r_{A,B}\gtrsim 5b/4$, 
and in principle we don't need to expand the phase (\ref{Psi}) in powers of $b/r$.
This expansion is still interesting though, in order to understand better the structure of the probability. 
For $r_A \rightarrow \infty$ and $r\equiv r_B$, we have then at the order $b^2/r^2$
\bea
    &&\mathcal{P}_{f\to f'} = \sin^2(2\Omega)\\
    &\times &\Bigg[\sin^2\Bigg( \frac{\Delta m^2}{8E_0} \Bigg[ 2r\left(1+\frac{r_S}{r}-\frac{r_Q^2}{2r^2} \right)
    - g(r)\Bigg] \Bigg) \nn
    &&~~~~\times \cos \left( \frac{m_1^2}{4E_0}f(r)\right) \cos \left( \frac{m_2^2}{4E_0}f(r)\right) \nn
    &&~~~~ +\sin^2 \left( \frac{\Delta m^2}{8E_0}f(r)\right)\sin^2 \left( \frac{\Sigma m^2}{8E_0}f(r)\right) \Bigg] ~,\nonumber
\eea
where 
\bea
&&\Delta b^2=b_1^2-b_2^2~~,~~\Sigma b^2=b_1^2+b_2^2\\
&&\Delta m^2=m_1^2 - m_2^2 ~~,~~ \Sigma m^2 = m_1^2 + m_2^2\nn
&&g(r)=\frac{\Sigma b^2}{2r}\left(1 - \frac{r_S}{r}\right) - l_+\nn 
&&f(r)=\frac{\Delta b^2}{2r}\left(1 -\frac{r_S}{r}\right) - l_-\nn
&&l_\pm=\frac{\pi}{2}\left(\frac{1}{b_1} \pm \frac{1}{b_2}\right)\left( \frac{3}{4}r_S^2 - r_Q^2\right)~.\nonumber
\eea
For a detection far from the compact object, the above probability reads
\bea
&&\lim_{r\to\infty}\mathcal{P}_{f\to f'}= \sin^2(2\Omega)\\
&\times&\Bigg[\sin^2\left(\frac{\Delta m^2}{8E_0}[2r_S+l_+]\right)\cos\left(\frac{m_1^2l_-}{4E_0}\right)\cos\left(\frac{m_2^2l_-}{4E_0}\right)\nn
&&~~~~~~~~~~~~~+\sin^2\left(\frac{\Delta m^2l_-}{8E_0}\right)\sin^2\left(\frac{\Sigma m^2l_-}{8E_0}\right)\Bigg]~,\nonumber
\eea
where $r$ is chosen such that $\Delta m^2 r/(4E_0)=2q\pi\to\infty$.
As one can see, the effect of the second order terms $r_S^2$ and $r_Q^2$, accumulated along the geodesic, leads to non-trivial contributions via the lengths $l_\pm$.

\subsection{Flavour interferometry}

In the weak-field regime $\epsilon\equiv r_S/b\ll1$ and for the purpose of visualisation, 
we will consider the asymptotically flat space, with Cartesian coordinates $(x,y)$ and $r=\sqrt{x^2+y^2}$.
The trajectories are then approximated by hyperbolas, with asymptotic straight lines defined by the deflection angle $\varphi$. 
This angle is calculated to the second order in $\epsilon$ for null geodesics in \cite{Briet:2008mz}
\be
\varphi=2\epsilon+\frac{3\pi}{4}\left(\frac{5}{4}-\frac{r_Q^2}{r_S^2}\right)\epsilon^2+{\cal O}(\epsilon^3)~,
\ee
and the equation of the asymptotic lines is 
\be
\frac{y-b}{x}\simeq-\tan(\varphi)~.
\ee
In the flat space approximation though, we will keep only the first order in $\epsilon$ for the angle $\varphi$.
The impact parameters of the trajectories corresponding to the intersection point $(x,y)$ are then given by
\bea
b_1&\simeq&\frac{y}{2}+\frac{1}{2}\sqrt{y^2+8xr_S}~~>~0\\
b_2&\simeq&\frac{y}{2}-\frac{1}{2}\sqrt{y^2+8xr_S}~~<~0~.\nonumber
\eea
We plot here the probability (\ref{proba}), with the following values of parameters 
\bea
&& m_1 = 10^{10}r_S^{-1} ~~,~~ m_2 = 0.99 \times 10^{10}r_S^{-1}  \\
&& E_0 = 10^7 \times m_1 ~~,~~\Omega = \frac{\pi}{4}~.\nonumber
\eea

\begin{figure}[htpb]
    \centering
    \includegraphics[width=0.48\textwidth]{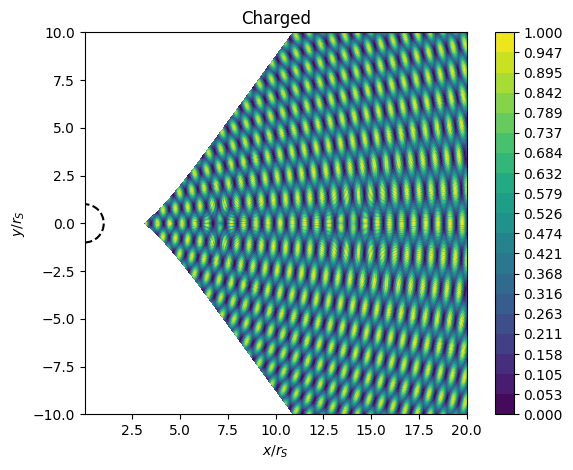}
    \caption{ Oscillation probability for a nearly maximal charge $r_Q=0.49r_S$ and a mass difference of $1\%$.
    \label{1_perc_mass_diff_maximal_charge}}
\end{figure}

\begin{figure}[htpb]
    \centering
    \includegraphics[width=0.48\textwidth]{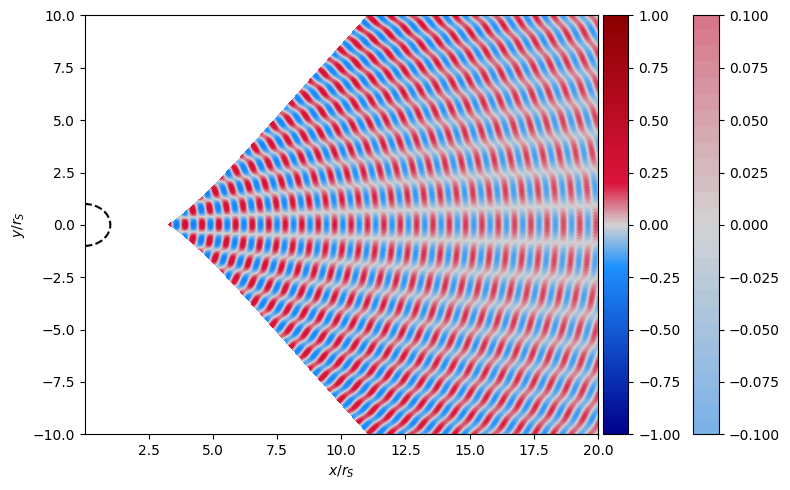}
    \caption{ Difference of oscillation probability $\Delta P_{f\to f'}$ (charged minus uncharged) for a mass difference of $1\%$.
    \label{1_perc_mass_diff_change_plot}}
\end{figure}

The event horizon ($r=r_S$) is marked by a dashed semi-circle, and the incident plane wave comes from the left. Furthermore, to be consistent with our approximations, only the region $r>5b/4$ is plotted. Fig. \ref{1_perc_mass_diff_maximal_charge} shows the interference pattern for a compact object of nearly maximal charge of $r_Q=0.49r_S$. Comparing the interference pattern with that of a Schwarzschild metric $r_Q=0$, we find a significant difference as shown in Fig. \ref{1_perc_mass_diff_change_plot}. The effect of the charge manifests as a shift of the interference pattern.

\begin{figure}[htpb]
    \centering
    \includegraphics[width=0.48\textwidth]{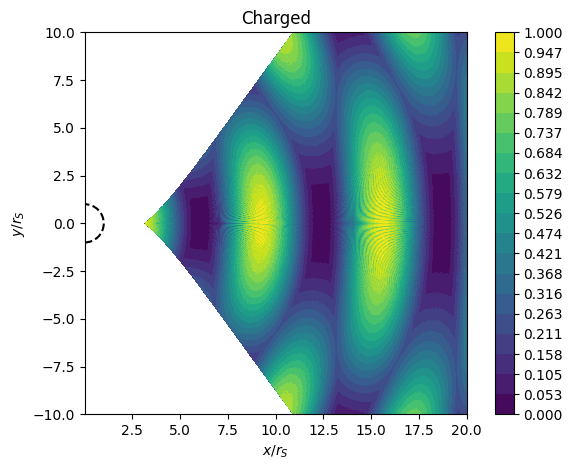}
    \caption{ Oscillation probability for a nearly maximal charge $r_Q=0.49r_S$ and a mass difference of $0.1\%$.
    \label{0.1_perc_mass_diff_maximal_charge}}
\end{figure}

\begin{figure}[htpb]
    \centering
    \includegraphics[width=0.48\textwidth]{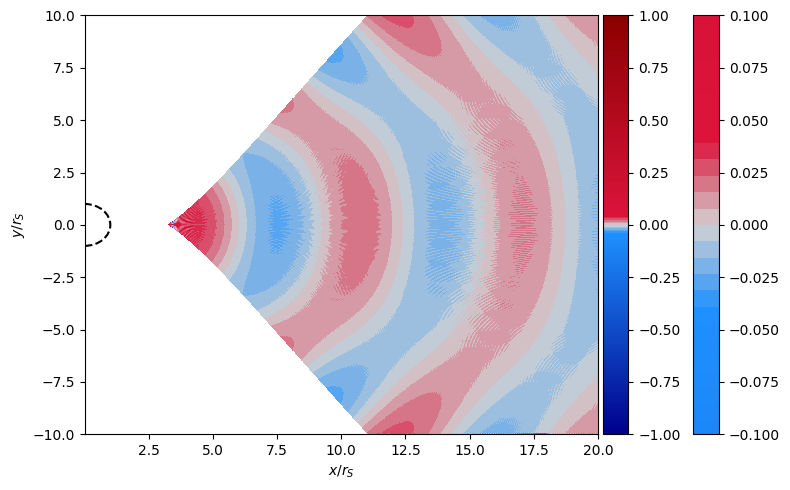}
    \caption{ Difference of oscillation probability $\Delta P_{f\to f'}$ (charged minus uncharged) for a mass difference of $0.1\%$.\label{0.1_perc_mass_diff_change_plot}}
\end{figure}

The interference pattern and the effect of the charge can be seen in more detail by considering a mass difference of $0.1\%$, where $m_2 = 0.999 \times 10^{10}r_S^{-1}$, as is done in Figs. \ref{0.1_perc_mass_diff_maximal_charge} and \ref{0.1_perc_mass_diff_change_plot}.

\begin{figure}[htpb]
    \centering
    \includegraphics[width=0.48\textwidth]{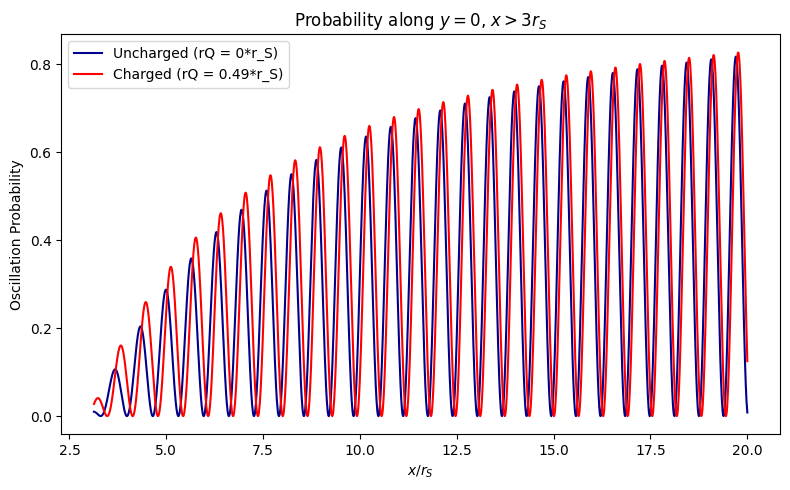}
    \caption{Comparison between the probability of oscillation caused by a black hole of charge $r_Q = 0.49 r_S$ and charge $r_Q =0$ along the axis $y=0$.\label{1_perc_mass_diff_1D_diff_y=0} }
\end{figure}

\begin{figure}[htpb]
    \centering
    \includegraphics[width=0.48\textwidth]{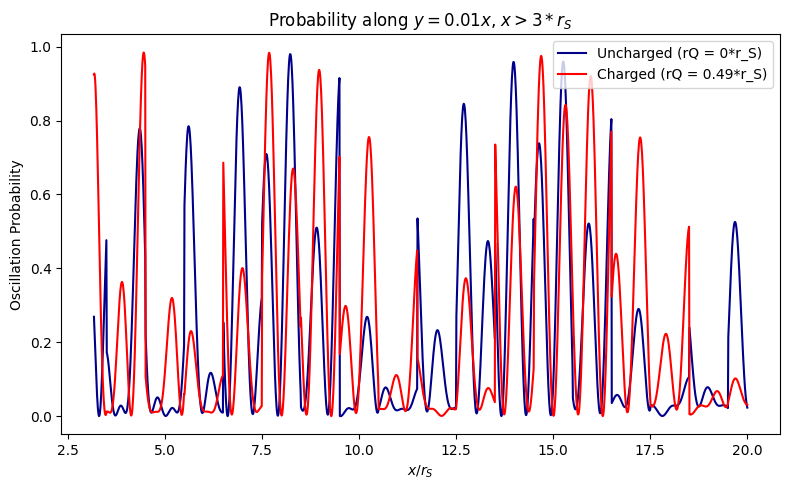}
    \caption{Comparison between the probability of oscillation caused by a black hole of charge $r_Q = 0.49 r_S$ and charge $r_Q =0$ along $y=0.01x$. \label{1_perc_mass_diff_1D_diff_y=0.01x} }
\end{figure}

It is also interesting to consider the oscillation probability along specific axes, which is done in Figs. \ref{1_perc_mass_diff_1D_diff_y=0} and \ref{1_perc_mass_diff_1D_diff_y=0.01x}, along the axis $y=0$ and $y=0.01x$ respectively. We note that we choose to start from $x>3 r_S$ since the weak field approximation becomes unreliable below this threshold.

\section{Conclusion}

The interference pattern for flavour oscillations, induced by a spherically symmetric gravitational background, 
leads to two different oscillation lengths: one relevant to the radial direction, which depends on the difference of masses squared, 
and the other relevant to the orthoradial direction, which depends on the sum of the masses squared. 
The measurement of this interference pattern is difficult to achieve, 
but it is worth keeping in mind that such a setup could lead to the individual values of the masses.
We found that the effect of a charge is not negligible since, compared to the Schwarzschild case, 
it significantly shifts the interference pattern.
This feature could lead to a better understanding of astrophysical events, 
particularly those emitting neutrinos subjected to gravitational lensing from a compact object.

From a technical point of view, we proposed a systematic way to treat artificial divergences, obtained as a result of a naive expansion of the phase in powers of the horizon radius over the impact parameter. This prescription is based on an analytical continuation, it reproduces known results at the first order in this expansion, and it is easy to apply to higher orders. 

We note that the results derived here are also valid for fermions, since spin has no effect on the oscillation mechanism. 
There is the additional phenomenon of spin flip for fermions though \cite{Dolan:2006vj}, 
although it doesn't happen in the ultra-relativistic regime \cite{Dvornikov:2021hps}, where the mass is negligible compared to the energy.

\section*{Acknowledgments}

This work was motivated by the Master thesis of EM, supervised at King's College London by JA. 
The authors would like to thank Katy Clough and Fannie Vaquero for helpful discussions.
The work of JA was supported by the United Kingdom Science and Technology Facilities Council (STFC) [Grant No.~ST/T000759/1].
For the purpose of open access, the authors have applied a Creative Commons Attribution (CC BY) licence to any Author Accepted Manuscript version arising.

\bibliography{interference_pattern}

\end{document}